\documentclass[twocolumn,twoside,slac]{revtex4}
\usepackage{moreverb}
\usepackage{graphicx}
\usepackage{fancyhdr}
\begin{document}

\title{Detector Description Framework in LHCb}

\author{Sébastien Ponce, Pere Mato Vila, Andrea Valassi}
\affiliation{CERN, Geneva, Switzerland}
\author{Ivan Belyaev}
\affiliation{ITEP, Moscow, Russia}
\affiliation{CERN, Geneva, Switzerland}

\begin{abstract}

The Gaudi architecture and framework are designed to provide a common
infrastructure and environment for simulation, filtering, reconstruction
and analysis applications. In this context, a Detector Description Service
was developed in LHCb in order to also provide easy and coherent access
to the description of the experimental apparatus.
This service centralizes every information about the detector,
including geometry, materials, alignment, calibration, structure and
controls. From the proof of concept given by the first functional
implementation of this service late 2000, the Detector Description Service
has grown and has become one of the major components of the LHCb
software, shared among all applications, including simulation, reconstruction,
analysis and visualization.

We describe here the full and functional implementation of the service.
We stress the easiness of customization and extension
of the detector description by the user, on the seamless integration
with condition databases in order to handle time dependent data
and on the choice of XML as a persistency back-end for LHCb Detector data.
We show how the detector description can be shared but still contains
application specific data and keeps at the same time several levels of
granularity.
We present several external tools which provide additional value to the
Detector Description Service like a dedicated, easy to use XML editor
and different geometry checkers.
We finally give hints on how this service could evolve to be part of
a common effort between all LHC experiments that would aim at defining
common Detector description tools at the level of the LCG project.

\end{abstract}

\maketitle

\thispagestyle{fancy}

\section{Introduction\label{SecIntro}}

This paper is a follow-up of a previous paper presented a CHEP'00
\cite{RadovanPaper}. There was presented a very first
implementation of the LHCb framework for Detector Description as well as ideas
for the future development of the project.

Since 2000, much progress was made and ideas have been changed into actual
software which is now in production. We thus want to describe here the general
structure of the final framework and insist on the new evolutions, particularly
on the different possibilities of extensions the user can use to adapt the
framework to her/his needs. We also describe how the framework integrates
with a condition database and the different tools provided to ease the access
and retrieval of detector data.

All these topics are discussed in the following sections :

{\bf Section \ref{SecBasics}} gives an outline of the basics of the detector
description framework, including the data it works on and the way they are
stored in both the transient and the persistent representation.

{\bf Section \ref{SecExtensions}} deals with the various extensions that a
user can make to the generic framework in order to adapt it to her/his own
needs. This includes the usage of generic parameters, the definition of new
C$^{++}$ objects and the extension of the data schema itself.

{\bf Section \ref{SecCondDB}} shows how a condition database can be easily
integrated to the framework in order to deal with time/version varying data.
It explains how the impact on the end user is reduced to the very minimal.

{\bf Section \ref{SecTools}} lists and quickly describes the different
tools provided to the end user to interface the framework. These allow to
easily edit data, check the described geometry and visualize it together
with event data.

\section{Detector Description Basics\label{SecBasics}}
The purpose of the detector description framework is to store, access and
process all data related to the description of the detector and used by
programs like detector simulation, event reconstruction, physics analysis,
visualization, ...

\subsection{Data Content\label{SubSecContent}}

\begin{figure*}[ht]
\begin{center}
\includegraphics[scale=0.4]{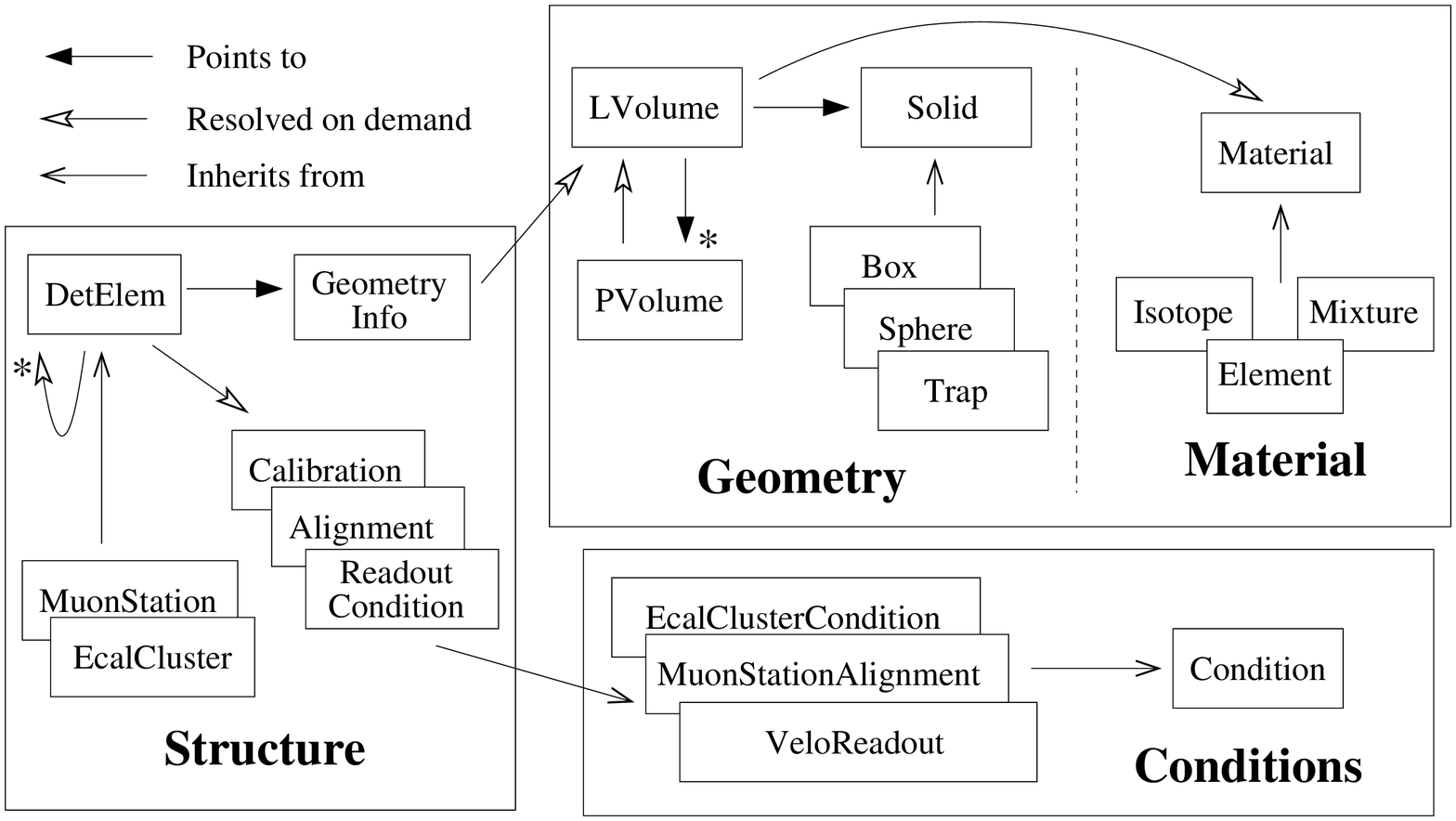}
\caption{LHCb Detector Description framework content.}
\label{FigDetDescStructure}
\end{center}
\end{figure*}

The description of the LHCb detector was split into three parts in the LHCb
framework, as depicted on figure~\ref{FigDetDescStructure} :
\begin{itemize}
\item{\bf Geometry and Materials} This first part contains the classical physical
description of the detector. This encompasses the geometry itself as well
as the materials of the detector and their properties. The description is
very similar to the Geant4 \cite{Geant4} approach, based on logical and
physical volumes so we will not detail it further here.
\item{\bf Structure of the detector} This part consists in
an abstract hierarchical description of the elements that compose the
detector. The basic element here is naturally called Detector Element and
acts as a center of information for all data dealing with the description
of a given subpart of the detector. Among others, a detector element has a
pointer to the related geometry, the calibration and alignment data and the
readout conditions. Note that detector elements can be extended and adapted
by the users as we'll describe in more details in section \ref{SecExtensions}.
\item{\bf The detector conditions} Conditions are extensible objects that
can contain time/version varying da\-ta. The extensibility is describe in
details in section \ref{SecExtensions} and is exactly identical to the
extensibility of detector elements. The specific part of the conditions
is actually the time/version variance. A given data stored in a condition
object may thus have different values depending on time but also on versions.
Examples of condition objects are data describing the alignement of the
detector, its calibration, the slow control parameters, ...
See section \ref{SecCondDB} for more details.
\end{itemize}

The steering part in all this is the structure part. Actually, every request
to the detector description framework has to start with the access to a Detector
Element. Other data can then be reached from there. It's worth noticing that
the granularity of the memory loading of information is actually finer than the
detector element itself, as show on figure~\ref{FigDetDescStructure}. Many
references in it are actually only loaded on demand. The detail mechanism
of the data access is described in detail in the next section.

\subsection{Transient Store mechanism\label{SubSecTransientStore}}

The access to the detector description data in a Gaudi application follows the
``transient store'' mechanism of the Gaudi framework
\cite{Gaudi00, Gaudi01, GaudiProject}.
The transient store is a place in memory where the detector elements used by the
application are loaded and cached. It is a hierarchical, tree like structure
where elements are designated by paths, as in Unix file-systems.

\begin{figure}[ht]
\begin{center}
\includegraphics[scale=.44]{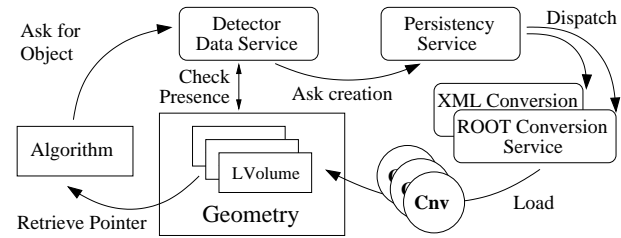}
\caption{Transient store mechanism in Gaudi.}
\label{FigTransientStore}
\end{center}
\end{figure}

The loading of detector elements into the transient store is only done on
demand and uses the mechanism described in figure \ref{FigTransientStore}.

Here is a sketch of what happens when a given data is accessed :
\begin{itemize}
\item all starts when an ``Algorithm'' asks the detector data service for a given
detector element.
\item the service will first look in the transient store whether this element
is already present. If it finds it, the pointer to the element is returned
directly.
\item if the element is not in the transient store, the detector data service
will request the persistency service to load it.
\item the persistency service, depending on the technology used to store the
data will ask the relevant conversion service to convert the data from the
given technology into C$++$ transient store objects. The conversion service
will actually use a set of converters specialized in the conversion of the
different possible objects to achieve this task.
\item finally, the detector data service can return a pointer to the new object
in the transient store.
\end{itemize}

\subsection{The persistent storage of detector data\label{SubSecPersistentStorage}}

As explained in the previous section, the Gaudi transient store mechanism
and the different services involved in the data loading allow an application
to be fairly independent of the technology used to store the data.
However, if one wants to use a given technology, she/he has to get/provide
the corresponding conversion service as well as a set of converters for
her/his object classes.

In the case of detector data in LHCb, we currently provide an XML conversion
service which is able to store/retrieve data from XML files and the
corresponding set of converters for detector description objects.
The choice of XML was driven mainly by its easiness of use and the number
of tools provided for its manipulation and parsing. Moreover, XML data can be
easily translated into many other format using tools like XSLT processors.

The grammar used for the LHCb XML data is pretty simple and straight forward,
actually very similar to other geometry description languages based on XML
like GDML \cite{GDML}. You can see on figure~\ref{FigXMLGeoCode} a short
example of a geometry description. However, the XML specification had some
limitations that we had to remove.

\begin{figure}[ht]
\begin{boxedverbatim}
<macro name="length" value="1*m"/>
<macro name="radius" value="20*cm"/>
<macro name="epsilon" value=".1*mm"/>
<subtraction name="sub2">
  <box name="mainBox"
       sizeX="length"
       sizeY="length"
       sizeZ="length"/>
  <tubs name="hole"
        outerRadius="radius"
        sizeZ="length+epsilon"/>
</subtraction>
<posXYZ z="length/2"/>
<rotXYZ rotX="90*degree"/> \end{boxedverbatim}
\caption{Sample of XML code describing the detector geometry.}
\label{FigXMLGeoCode}
\end{figure}

A first enhancement was to deal with the lack of types in XML data.
As a matter of fact, in pure XML, everything is a string.
Thus if you specify a length, you will get the string ``1*m''
instead of a numerical value. In order to solve this problem, the Gaudi XmlSvc
allows any string to be evaluated via the CLHEP expression evaluator
\cite{CLHEPExpressionEvaluator}. It also allows the user to define macros
that can be reused in expressions, as shown in the code of
figure~\ref{FigXMLGeoCode}.

Another big limitation is the fact that XML data cannot be
split among several XML files. In other words, there is no way to make
cross references among different XML files. We thus added a small extension
to the basic XML specification which allows this. The rule is that every
XML tag with a name ending with ``ref'' is actually a reference to another
tag with the same name, without the ``ref''. The syntax of the reference
itself is very simple : a URL containing the file name and the name of
the pointed tag.

\begin{figure}[ht]
{\bf Main.xml}
\begin{boxedverbatim}
<detelem name="Main">
  <geometryinfo lvname="/dd/Geometry/lvMain">
  <detelemref href="External.xml#Beam"/>
  <detelemref href="External.xml#Base"/>
  <detelemref href="Detector.xml#Velo"/>
  <detelemref href="Detector.xml#Rich"/>
</detelem>
\end{boxedverbatim}
\vspace{2mm}
{\bf External.xml}
\begin{boxedverbatim}
<detelem name="Beam">
  <geometryinfo lvname="/dd/Geometry/lvBeam"
                support="/dd/Structure/Main"/>
</detelem>
<detelem name="Base">
  <geometryinfo lvname="/dd/Geometry/lvBase"
                support="/dd/Structure/Main"/>
</detelem>
\end{boxedverbatim}
\caption{Reference mechanism in XML.}
\label{FigXMLRefs}
\end{figure}

This reference mechanism allows to split the description of a hierarchical
system as a detector among several files so that each subpart of the system
is described in a different set of files, handled and edited by the
appropriate person. You can see a small example of this in figure
\ref{FigXMLRefs} where a detector structure is defined.

\section{Extensions of the default schema\label{SecExtensions}}

The schema presented in section \ref{SecBasics} allows to describe standard
detector geometry, materials and structure. However, this is not sufficient
for the vast majority of real life examples since we cannot extend the
default objects in order to include more specific parameters.

We address this issue by providing three possibilities of extension of the
default schema. Depending on the degree of flexibility required, the extension
will of course require more or less work from the user. The following
subsections describe the details of the extension process.

\subsection{Parameter extension\label{SubSecParams}}

This is the most simple but also most limited extension. It allows the user
to add parameters to any detector element or condition. As show on figure
\ref{FigXMLParameter}, a parameter is simply a triplet (Name, Value, Type).
The type of the parameter can be int, double or string.

\begin{figure}[ht]
\begin{boxedverbatim}
<detelem name="MStation01">
  <param name="Al_thickness"
         type="double">
    1.2222*mm
  </param>
</detelem>\end{boxedverbatim}
\caption{Definition of a parameter in XML.}
\label{FigXMLParameter}
\end{figure}

The parameters can then be very easily accessed in the C$^{++}$ framework as
show on figure \ref{FigCppParameter} using the {\it param} method.

\begin{figure}[ht]
\begin{boxedverbatim}
SmartDataPointer<IDetectorElement> station
  (detSvc(),
   "/dd/Structure/LHCb/Muon/MStation01");
std::cout << station->param("Al_thickness");\end{boxedverbatim}
\caption{Access to a parameter in C$^{++}$.}
\label{FigCppParameter}
\end{figure}

This first extension possibility allows the user to store any extra data in the
detector ele\-ments\-/\-con\-di\-tions and to retrieve them. However, these data are
not structured and cannot be used to extend the behavior of the \verb#DetectorElement#
and \verb#Condition# objects.

\subsection{C$^{++}$ class extension\label{SubSecCppExt}}

The second extension possibility concerns the C$^{++}$ objects.
As permitted by the C$^{++}$ language itself, one can extend and customize
the default objects \verb#DetectorElement# and \verb#Condition# by simply
inheriting from them. This allows addition of any new member or method
needed to describe the specific behavior of a given detector subpart.

A special method called \verb#initialize# is also provided for customization
of the object creation. This method is actually called just after the creation
of any new object of the given type by the framework.

Figure~\ref{FigClassExt} gives an idea of what a class extension can provide.
Note the use of a parameter in this example for the initialization of a new
member. The different extension mechanisms are of course not exclusive.

\begin{figure}[ht]
\begin{boxedverbatim}
class MyDetElem : public DetectorElement {
public:
  const CLID& clID() { return classID(); }
  static const CLID& classID() { return 999; }
  int getChannelNb() { return chNb; }
  StatusCode initialize() {
    chNb = paramAsInt ("ChNb");
    return SUCCESS;
  }
private:
  int chNb;
}
\end{boxedverbatim}
\caption{Extension of the default DetectorElement class in C$^{++}$.}
\label{FigClassExt}
\end{figure}

So far so good but one may ask how the framework is aware of the existence
of new objects and can create them. This is done by associating a unique ID
to any new object type (see figure~\ref{FigClassExt}) and mentioning it in the
XML code (see XML code in figure~\ref{FigDTDExt}, the mention of an ID is
independent of the DTD extension).

Besides the mention of the specific ID of the new object, one should provide
the framework with a specific converter for this new object type (see
subsection~\ref{SubSecTransientStore} for the conversion mechanism details).
This step is highly simplified, and almost automatized by the existence of
a default templated converter that does everything needed for the user.
One still need to actually declare the new converter by extending the default
one. This is the first line of code in figure~\ref{FigCnvDecl}.
The two other lines deal with the instantiation of the new converter by the
framework using the Gaudi standard abstract factory pattern.

\begin{figure}[ht]
\begin{boxedverbatim}
typedef  XmlUserDetElemCnv<MyDetElem>  MyCnv;
static CnvFactory<MyCnv> MyFactory;
const ICnvFactory& XmlMyFactory = MyFactory;
\end{boxedverbatim}
\caption{Simple converter implementation}
\label{FigCnvDecl}
\end{figure}

\subsection{Full extension\label{SubSecFullExt}}

The C$^{++}$ class extension combined with the parameters allowed to add data to
the default schema and to specialize the behavior of the \verb#DetectorElement#
and \verb#Condition# objects. However, there is still no way to add structured
data and to map them easily to structured members in the C$^{++}$ world.
This is what we provide here with the so-called ``full extension''.

Here are the new possibilities :
\begin{itemize}
\item extend the XML DTD to define new tags and attributes
\item extend the C$^{++}$ objects as before
\item map the new XML tags/attribute to the new objects members via the writing of
dedicated converters
\end{itemize}

The second step was the subject of the previous subsection.
Let's detail the first and last step.

\begin{figure}[ht]
\begin{boxedverbatim}
<?xml version="1.0" encoding="UTF-8"?>
<!DOCTYPE DDDB SYSTEM "../DTD/structure.dtd" [
<!ELEMENT channelSet (channels*)>
<!ELEMENT channel EMPTY>
<!ATTLIST channelSet name CDATA #REQUIRED>
<!ATTLIST channels name CDATA #REQUIRED
                   nb   CDATA #REQUIRED>
]>
<detelem name="Head" classID="1234">
  <specific>
    <channelSet name="Controls">
      <channels name="in" nb="20"/>
      <channels name="out" nb="150"/>
    </channelSet>
  </specific>
</detelem>
\end{boxedverbatim}
\caption{Extension of the geometry description language.}
\label{FigDTDExt}
\end{figure}

\subsubsection{Extension of the XML DTD\label{SubSubSecXMLExt}}

The extension of the new DTD by the user almost is completely unconstrained.
The only limitation is that the new DTD tags must be children of the
already defined \verb#<specific># tag of the default DTD. Figure
\ref{FigDTDExt} shows an example of a simple extension of the DTD. We
will not go into more details on this since it is standard XML code.

\subsubsection{Writing a new converter\label{SubSubSecWritingConv}}

In order to parse the new DTD elements and retrieve the corresponding data
from XML, the user has to write a specialized converter. As in the case of
the extension of C$^{++}$ objects, she/he can reuse the default converters
by inheriting from them.

\begin{figure}[ht]
\begin{boxedverbatim}
StatusCode XmlMyCnv::i_fillSpecificObj
           (DOM_Element e, MyDetElem* d) {
  if ("ChannelSet" == e.getNodeName()) {
    DOM_NodeList list = e.getChildNodes();
    int nbTot = 0;
    for (int i=0; i<list.getLength(); i++) {
      DOM_Node c = list.item(i);
      DOM_Element ce = (DOM_Element &) c;
      if ("channels" == ce.getNodeName())
        nbTot += xmlSvc().eval
          (ce->getAttribute("nb"));
    }
    d->setChannelNb(nbTot);
  }
}
\end{boxedverbatim}
\caption{Writing a dedicated converter in C$^{++}$.}
\label{FigConvExt}
\end{figure}

But this time, she/he has to provide some real
code in order to parse the new DTD elements. This code has to use the
Xerces XML parser \cite{Xerces} as shown in the example on figure
\ref{FigConvExt}. It simply consist in reimplementing a single method
which is called for every child of a \verb#<specific># node.
Writing such code should be rather straightforward, even for users that
are not used to XML.

\section{Condition database\label{SecCondDB}}

The previous sections described how to store, retrieve and customize detector
related data in a very static way. By static we mean that each parameter can
only have one single value whatever the time and the version of the detector
on deals with. This is but not the case in real life. The most obvious example
is the temperature of a given element of the detector. Moreover, the structure
of the detector itself may vary over time. As an example, the Velo sub-detector
of LHCb can be placed in two different positions depending on the beam conditions.

We describe in this section how we address this issue by integrating a condition
database into the framework.

\subsection{Purpose and content of the database\label{SubSecPurpose}}

\begin{figure*}[ht]
\begin{center}
\includegraphics[scale=.65]{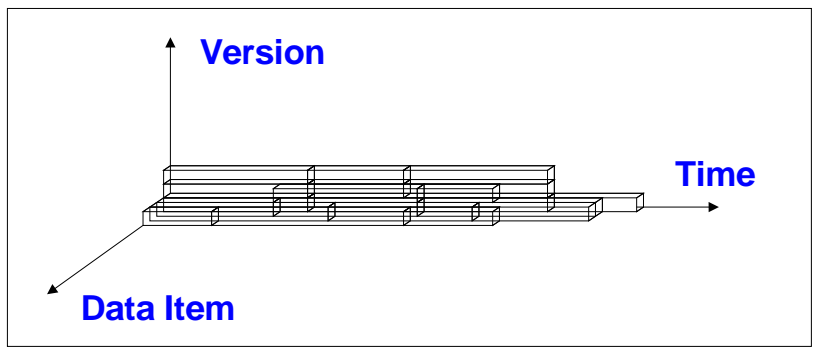}
\caption{Condition Database dimensions.}
\label{FigCondDB}
\end{center}
\end{figure*}

A condition database is a three dimensional da\-ta\-ba\-se, as depicted on
figure~\ref{FigCondDB} which contains all varying data. The three dimensions
are the data item, the time and the version of the detector. Each data item may
then have different values for different times and different versions of the
detector. One may argue that in real life, only one version of the detector
exists at a given moment in time and that the version dimension is mainly
useless. This is perfectly true for geometry. However, in the case of
calibration data, one could very well define different versions of the
calibration constants at the same place in time, and compare results of
the physics analysis in both cases.

\subsection{Implementation of the database\label{SubSecImpl}}

The implementation of the condition database is actually segmented into three
layers.
\begin{itemize}
\item An abstract interface defining the public interface to the database
\item The database itself
\item A Gaudi service using the abstract interface
\end{itemize}

\subsubsection{The condition database interface\label{SubSubSecAbsInt}}

The interface to the condition database was specified before the
implementation and agreed between the CERN IT division and all LHC
experiments \cite{CondDbInterfaceProp, CondDbInterface}.
Among other things, it specifies which data can be stored in the condition
database. These are chunks of bytes seen by the interface as char*
(in the C$^{++}$ world).

The actual purpose of the interface is to allow the use of different
implementations (with different databases in the backend) without changing
the Gaudi service code. We are thus currently able to store our condition
data into three different databases, namely Objectivity, ORACLE and MySQL.

\subsubsection{The Database\label{SubSubSecDB}}

Several implementations of the interface are available. Two of them
were provided by the CERN IT division, one using Objectivity \cite{CondDbObjty}
and the other using ORACLE as a backend.

A third implementation was provided recently by the Lisbon ATLAS group
\cite{CondDbMySQL} and uses the open-source database MySQL as a backend.
All of these implementations can be easily used by the LHCb software.
Currently, most of the data are stored using ORACLE. However, we plan
to use MySQL for local replicas of the data (e.g. on laptops).

\subsubsection{The Gaudi Service\label{SubSubSecGaudiSvc}}

In the pure spirit of Gaudi, the condition database was interfaced in LHCb as
a new service accessible via a well defined interface. This service
is completely independent of the actual implementation of the database since
it uses the condition database interface as described in the previous
section.

\begin{figure*}[!bt]
\begin{center}
\includegraphics[scale=.45]{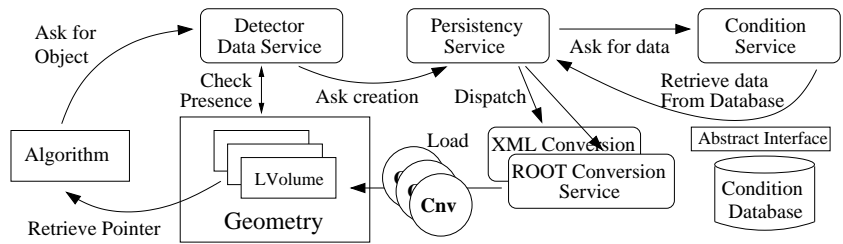}
\caption{Transient store mechanism in Gaudi using Condition database.}
\label{FigTransientStoreCondDB}
\end{center}
\end{figure*}

Actually, the service is also independent of the objects that we store in the database
since it is only used to get chunks of bytes that will be interpreted afterward by
the usual conversion service and its set of converters. Figure \ref{FigTransientStoreCondDB}
shows the sketch of an access to data when using the condition database.
\begin{itemize}
\item an algorithm first request some data to the Detector Data Service.
\item if the data already reside in the transient store a pointer to them is
returned. Note at this point that the data cannot be outdated and cannot use
a bad version of the detector. This is ensured by a proper cleanup of the
transient store each time the time or the version of the detector changes.
\item if the data are not present in the transient store, the persistency service
is called. It will realize that the data are stored in the condition database
and use the condition service to get them.
\item the condition service connects to the database via the interface
and retrieve data as a chunk of bytes.
\item this chunk is then given to the right conversion service that will use its
set of converters to convert it into C$^{++}$ objects.
\item the newly created object is then stored in the transient store and sent back
to the algorithm.
\end{itemize}

This sketch can be compared to the case described in
subsection~\ref{SubSecTransientStore} where no condition database is present.
You can see that the only addition is the retrieval of data through the
condition service instead of simple file reading.

\subsubsection{Impact on end user\label{SubSubSecImpact}}

The impact of the existence of condition data in the detector description is
actually very small from the point of view of the end user.
This is especially true for the access to the data, which still takes place
exactly as before and as shown on figure \ref{FigCppParameter}. Note than
the user still have to specify which version of the detector he wants to use.

\begin{figure}[ht]
\begin{boxedverbatim}
<DDDB>
  <catalog name="HCal">
    <!-- Hcal slow control catalog -->
    <conditionref href=
"conddb:/CONDDB/SlowControl/Hcal/scHcal#scHcal"
    />
  </catalog>
</DDDB>
\end{boxedverbatim}
\caption{Extension of the geometry description language.}
\label{FigCondRef}
\end{figure}

Concerning the data storage, there are some slight changes in order to allow the
framework to know when data are stored in the condition database instead of
regular XML files. Figure \ref{FigCondRef} shows the new XML code. The only
difference lies in the reference to the data which now uses a special
protocol named ``conddb''. The rest of the URL defines where the data
is located in the database.

\section{Tools\label{SecTools}}

We describe in this section the set of tools that are provided in order to
facilitate the usage of the detector description framework. This goes from
data edition to geometry checkers and geometry visualization.

\subsection{Data Edition\label{SubSecDataEdition}}

\begin{figure*}[ht]
\begin{center}
\includegraphics[scale=.45]{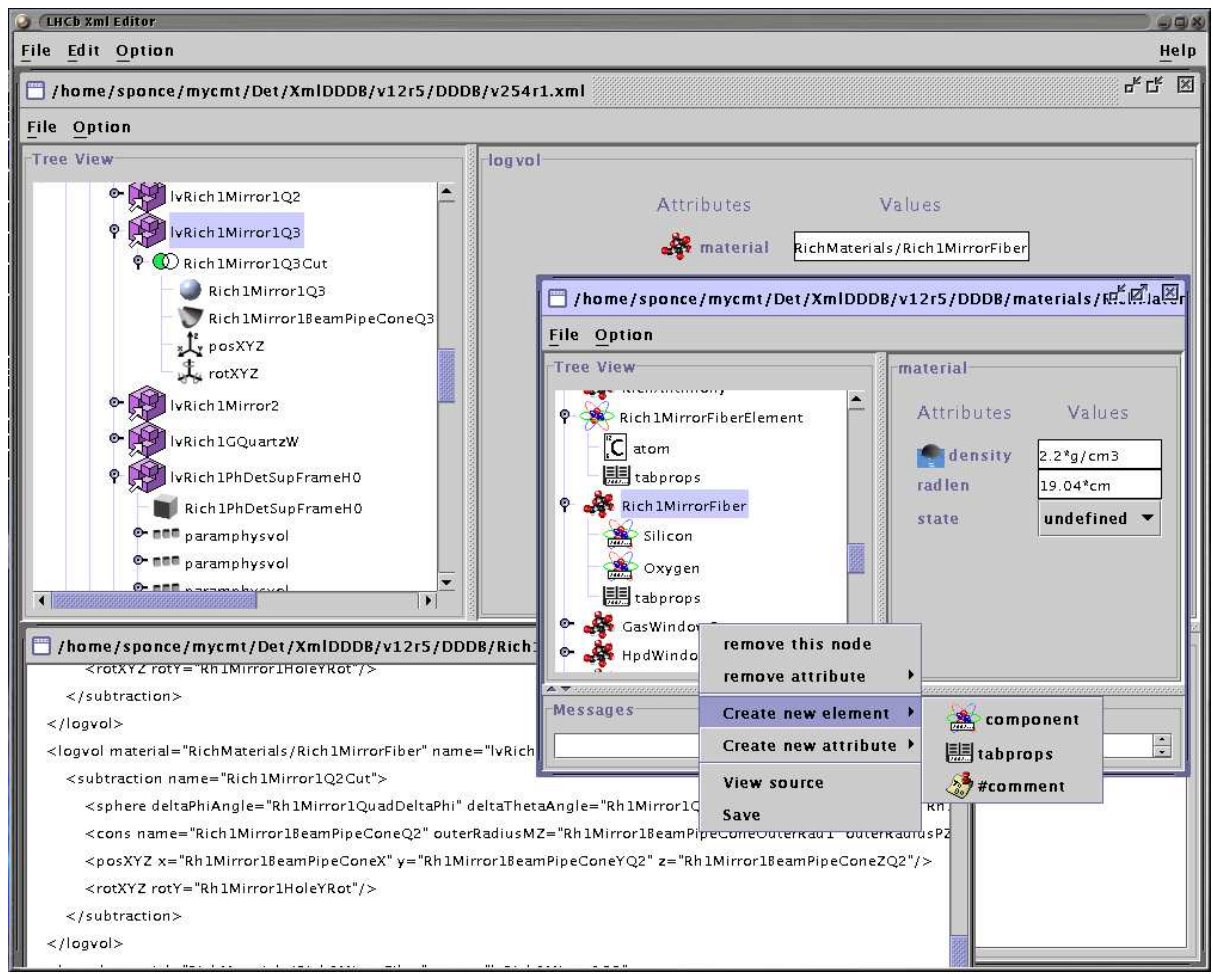}
\caption{XmlEditor graphical interface.}
\label{FigXMLEditor}
\end{center}
\end{figure*}

The main tool we provide for data edition is a specialized, graphical XML editor.
Its interface is shown on figure \ref{FigXMLEditor}. As you can see, it is
an explorer like, easy to use editor which incorporates all common features
like cut and paste, drag and drop of nodes, etc...

Its particularity compared to regular XML editor that one can find on the web
is to understand our small extension to XML dealing with references. In other
words, you can edit your data in this editor across files without bothering
with the references. The XML that you see and edit actually goes across files and
the only change for you is that he icon of the node you are editing has a small
arrow in a corner, indicating that you are editing the target of a reference.

Besides this, the tool was implemented in order that the XML code generated is
still easily readable by human beings (as an example, it is correctly indented).
It can thus be edited in parallel using more usual tools like (X)Emacs.

\subsection{Geometry checking\label{SubSecGeoCheck}}

One of the main problems arising when describing the geometry of a detector
is to validate it. Usually the validation is done in two steps : the first step
consists in looking at the geometry and correcting obvious errors. The second
step consists in verifying that there are no volume intersections in the geometry
that would prevent the simulation to work correctly. This intersections, when
small and hidden are actually very hard to detect by visual inspection.

The first step is achieved using the visualization software presented in
subsection~\ref{SubSecVisu}.
In order to achieve the second step, two main tools are provided,
a visual geometry checker and the transport service.

\subsubsection{Visual geometry checker\label{SubSubSecVisualCheck}}

The Visual geometry checker is based on the David tool \cite{David} :
the Dawn's Visual Intersection Debugger provided with the Geant4
\cite{Geant4} simulation package. It was interfaced to Gaudi using the Giga
package (``Gaudi Interface to Geant Application'') \cite{WiteksTalk, Giga}.

This tool allows to visually debug the geometry by making every intersection
appear in red on the screen. It provides a very good first pass debugging
but does not actually insure that the corrected geometry is error free.
From time to time, it even finds fake errors.

The problem of this tool is that it is fully based on graphical visualization.
This makes it very handy but also makes it suffer from the approximations that
are done in the computation of the graphical display of the geometry.
In order to insure error free geometry, one has to provide another tool,
free from these geometry display approximation and only dealing with
original geometry data.

\subsubsection{Transport Service\label{SubSubSecTransportSvc}}

The transport service tool is not initially a geometry correction tool. As
its name mention it, it provides services for propagating particles in a
given geometry.

The key point is that this transportation mechanism is very sensible to
every intersection of two volumes in the geometry description. This allowed
to change it into a powerful geometry checker. The principle is to make
many particles go through the geometry randomly so that the probability
to cross every volume is very high. If ever an intersection of volumes exists,
it will be found when crossing it and reported to the user.

This tool has proved to be more reliable and more precise that David.
However, it is not graphical and less handy.

\subsection{Visualization\label{SubSecVisu}}

The 3D visualization tool provided in the LHCb detector description framework
is called Panoramix\cite{Panoramix} and is actually far more than just a
geometry visualization tool. As a matter of fact, it is also able to
display event data, to deal with histograms and even to launch interactive
analysis of data through its python interface to Gaudi.

\begin{figure*}[ht]
\begin{center}
\includegraphics[scale=.37]{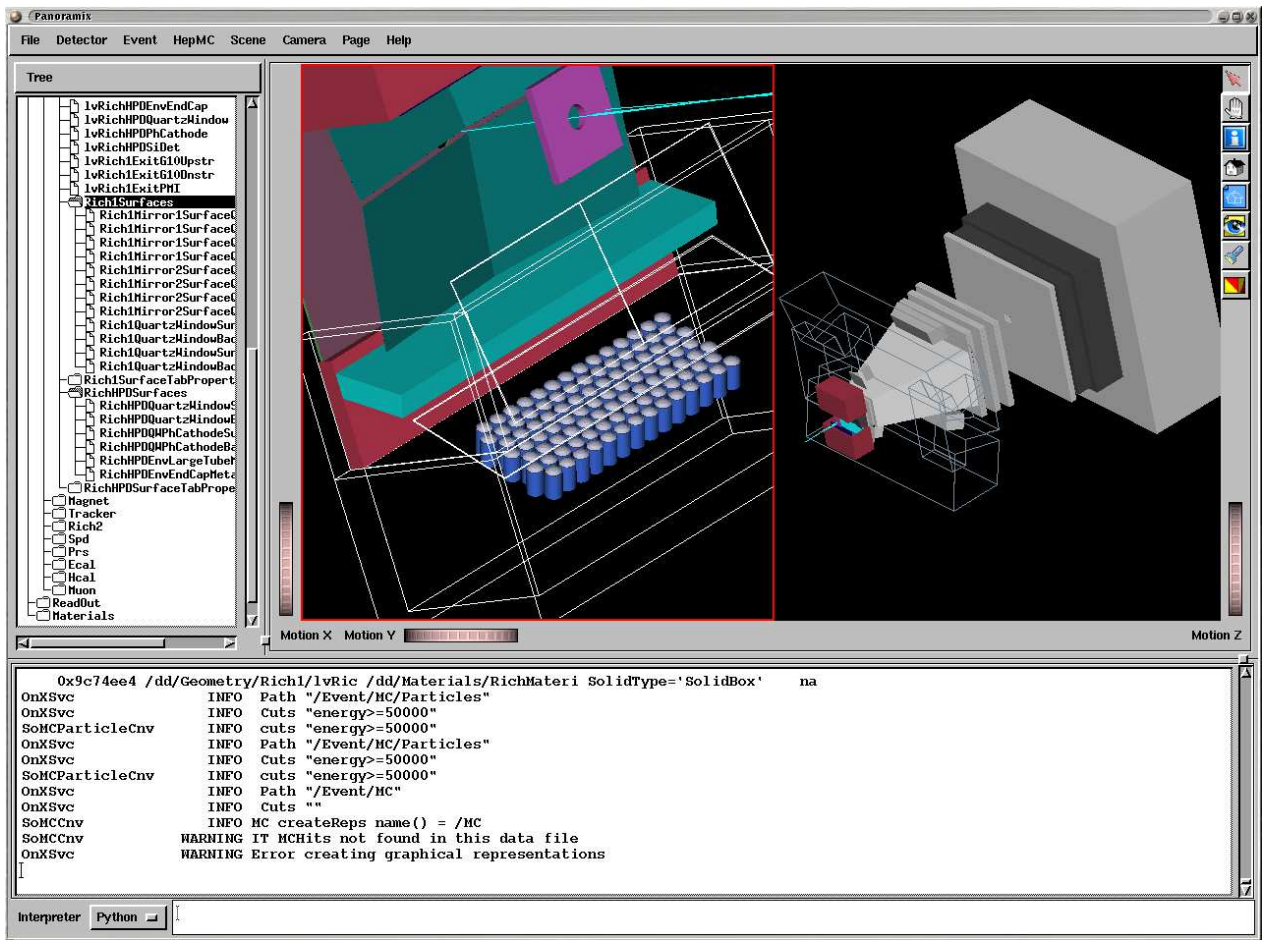}
\caption{Panoramix Visualization software.}
\label{FigPanoramix}
\end{center}
\end{figure*}

Figure \ref{FigPanoramix} shows a picture of the Panoramix user interface
as well as some pictures of the LHCb geometry.

\section{Conclusion\label{SecConclusion}}

The LHCb detector description framework presented here is nowadays fully
functional and proved in the past months to be stable and efficient.
It was actually used for our last data challenge and more than 50 millions
of events were reconstructed and analyzed using it.

The first simulations using Geant4 were also run showing that the sharing
of the detector description between Simulation, Reconstruction and Analysis
software works as expected.

\end{document}